\documentclass[modern]{aastex61}

\received{}
\revised{July, 2019}
\accepted{\today}
\submitjournal{PASP}

\shorttitle{Default Parallel Observations with JWST}
\shortauthors{Holwerda et al.}

\begin{document}

\title{Default Parallels: The Science Potential of JWST parallel Observations During TSO Primary Observations}

\correspondingauthor{Benne Holwerda twitter: @bennewillemholwerda}
\email{benne.holwerda@louisville.edu}

\author[0000-0002-4884-6756]{B.W. Holwerda}
\affiliation{Department of Physics and Astronomy, 102 Natural Science Building, University of Louisville, Louisville KY 40292, USA} 

\author[0000-0003-0910-5805]{Jonathan Fraine} 
\affiliation{Space Telescope Science Institute, 3700 San Martin Drive, 21218 MD, Baltimore, USA} 
\affiliation{Space Science Institute, 4750 Walnut St, Suite 205 Boulder, CO 80301, USA} 

\author[0000-0003-1609-5625]{Nelly Mouawad} 
\affiliation{Lebanese American University, 211 East 46th Street, New York, N.Y. 10017, USA} 

\author[0000-0002-8584-1903]{Joanna S. Bridge} 
\affiliation{Department of Physics and Astronomy, 102 Natural Science Building, University of Louisville, Louisville KY 40292, USA}


\begin{abstract}
The \emph{James Webb Space Telescope} (\emph{JWST}) will observe several stars for long cumulative durations while pursuing exoplanets as primary science targets for both Guaranteed Time Observations (GTO) and very likely General Observer (GO) programs. 
Here we argue in favor of an automatic default parallels program to observe e.g., using the F200W/F277W filters or grism of NIRCAM/NIRISS in order to find high redshift ($z>>10$) galaxies, cool red/brown dwarf- sub-stellar objects, Solar System objects, and observations of serendipitous planetary transits. 
We argue here the need for automated exploratory astrophysical observations with unused \emph{JWST} instruments during these long duration exoplanet observations.

Randomized fields that are observed in parallel mode reduce errors due to cosmic variance more effectively than single continuous fields of a typical wedding cake observing strategy \citep{Trenti08}. Hence, we argue that the proposed automated survey will explore a unique and rich discovery space in high redshift Universe, Galactic structure, and Solar System.

We show that the GTO and highly-probable GO target list of exoplanets covers the Galactic disk/halo and high redshift Universe, mostly well out of the plane of the disk of the Milky Way. Exposure times are of the order of the CEERS GTO medium deep survey in a single filter, comparable to CANDELS in \emph{HST}'s surveys and deep fields. The area covered by NIRISS and NIRCam combined could accumulate to a half square degree surveyed. 

\end{abstract}

\keywords{}

\section{Introduction} \label{sec:intro}



\begin{figure}
\begin{center}
	\includegraphics[width=\textwidth]{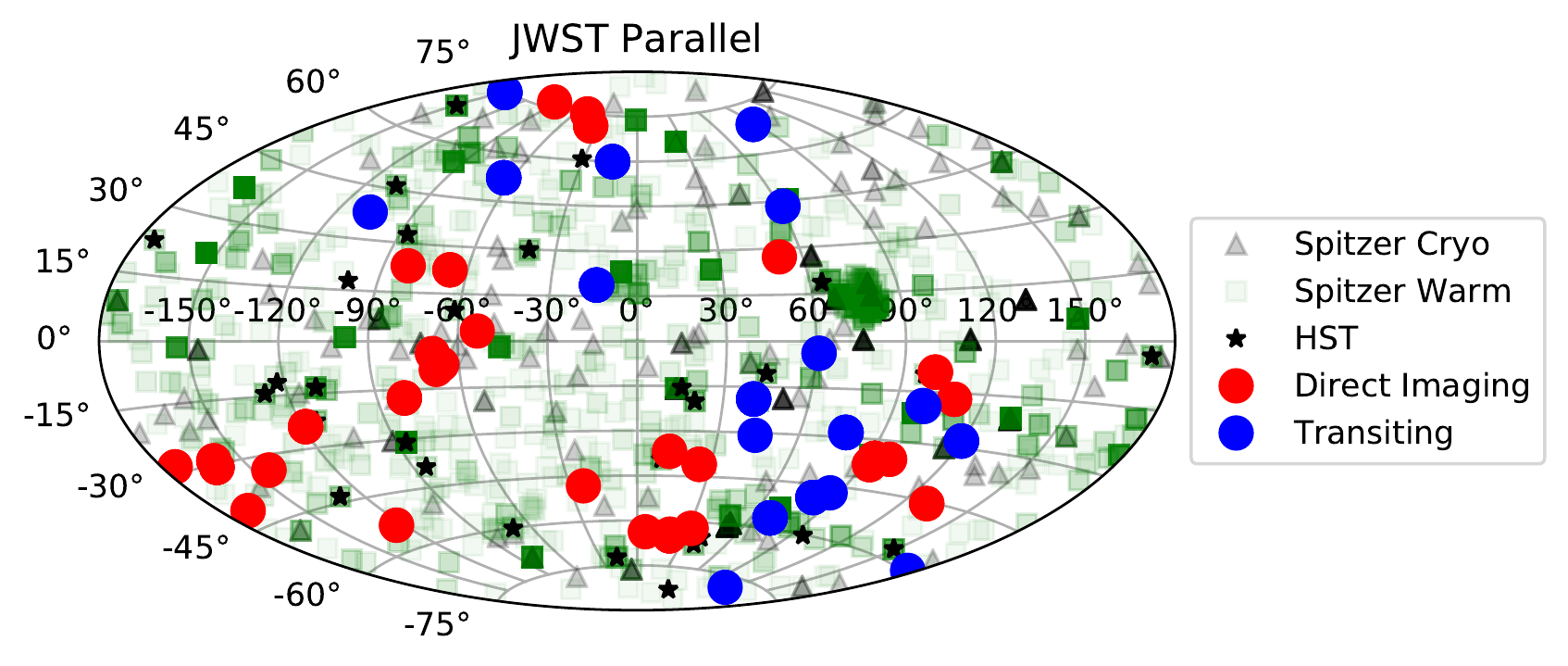}
	\caption{\label{f:allsky} The \emph{Spitzer} Cryo and Warm mission transient targets and \emph{HST} GO exoplanet programs (likely targets for \emph{JWST} GO follow-up proposals) and \emph{JWST} exoplanet GTO programs projected on the sky in Galactic Coordinates.  While there are multiple programs on a variety of targets, the number of pointings is $\sim30$ for the GTO (blue and red dots) and an unknown for the GO programs (depending on the GO success rate).  }
\end{center}
\end{figure}

Given the high price tag and the limited life expectancy of the \emph{James Webb Space Telescope} (\emph{JWST}), the pressure is on the astronomical community to maximize this new flagship's efficiency. Two observational communities that will make maximum use of this observatory and its revolutionary instrument suite are the exoplanet and the high-redshift galaxy population communities. Exoplanet direct imaging and transit studies require long, single-pointing exposures. Meanwhile, the other instruments on \emph{JWST} will be left unused for science. Given reasonable use (i.e., no filter changes, long integrations to limit buffer use), these can be put to excellent use for Solar System, Galactic, and extragalactic science (Figure \ref{f:allsky}). 

In order to minimize resources allocated to this program, we propose a single strategy to simplify the generation of this legacy data-set, ``default parallels".

default parallels would be planned as follows:
\begin{itemize}
\item Automatically examine in each non-primary instrument's FOV
\item If there is a bright target that would saturate deep photometry, then perform a transit observation (i.e., a time series of imaging with repeated short integrations).
\item If there are no bright sources, then do a deep (extragalactic) observation in a band redder than $H$, or using grism observations.
\item If the number of intermezzos is above three or four, deep imaging with multiple filters is performed.
\item If not enough principal observation intermezzos are available for deep imaging, grism observations are performed. 
\end{itemize}

The possibility of default parallels would have to also be calculated with the boundary conditions of no or few filter changes, no dithering, long integrations, and within \emph{JWST}'s limited onboard data storage space. 
Given the existing exoplanet \emph{JWST} GTO, ERS, and prospective GO observational expectations, it is possible that an default parallels program could observe up to a {\em half a square degree} -- accumulated area -- for deep galactic fields ($m<$28) in either F277W or F200W filters from NIRISS and NIRCam, spanning a large range sky coverage; these filters were selected because they are available on both instruments. We highlight double NIRISS and NIRCam observations as these can be homogeneous, wide field (2$\times$ (3'$\times$3') FOV) with limiting magnitudes ranging from $m_{AB} \sim 27-29$ mag (depending on actual in-flight performance etc).

MIRI observations (FOV 74" $\times$ 113") in default parallels would provide a separate unique, long wavelength coverage not available with any NIR instruments.
Below are a few science cases for such observations.



\section{default parallel Science Cases}

\subsection{Science Case A: High-redshift galaxies}

Long-stare NIRcam or NIRISS F200W/F277W observations would be beyond the $H$-band of \emph{Euclid} or the F184 filter on \emph{WFIRST}, or \emph{Hubble}'s F160W filter. These images will effectively  search for $H$-band dropout galaxies ($z>11$). Without ancillary data, these observations would be of limited immediate use for high-redshift studies, but with solid $H$-band fluxes, this could be an excellent constraint on the luminosity function of $z>11$ galaxy populations using the Lyman break technique \citep{Steidel96}. Critical here is that the $H$-band observations can be a follow-up of these \emph{JWST} parallel fields. The value of these preliminary search observations is a key investment for the discovery space of future follow-up observations. 

A similar program of \emph{HST}/WFC3 default parallels has been extremely successful in constraining the $z\sim8$ and $z\sim9$ brighter populations \citep{Trenti11,Trenti12,Trenti14a,Bradley12,Calvi16,Bernard16,Livermore18,Morishita18}, taking statistical power from the essentially random pointings, and therefore sampling a large volume \citep{Trenti08}. The random sampling also resulted in useful lower redshift ($z\sim 2$) clustering constraints \citep{Cameron19}, which would have been difficult to obtain otherwise.
Similar searches can be done with just \emph{JWST} F200W/F277W observations using NIRISS/NIRCam, supplemented with \emph{WFIRST/EUCLID/HST} near-infrared ($J$- and $H$-band) follow-up. 

During exoplanet phase curve observations and high-contrast imaging programs, there are nominally re-pointings to change filters (direct imaging) or data downloads (phase curves). During these significantly longer exposures, with sufficiently long intermezzos for re-pointings or data downloads, multiple filter observations can be considered for the non-primary instrument's default parallels, for example during transiting observations which consist of the greatest time commitments.
{We assume that during sensitive the exoplanet observations, a filter change on another instrument would be too disruptive and these should be reserved for breaks in the primary observation. If no break is available, the parallel would be a single, F200W or F277W deep image, with the longest on-ramp integrations for minimal storage use.
Multiple filter parallel observations --i.e. with primary observation breaks for filter changes in the parallel--} would expand the search space to $z>>11$ sources. These would be of almost similar quality as dedicated deep fields, without the specifically optimized dithering strategy however.

\subsection{Science Case B: Brown Dwarf Population of the Milky Way} 

Similarly, a search for the lowest-mass sub-stellar objects belonging to the Milky Way can benefit from observations such as the above default parallels. The randomized sampling of the Milky Way volume equally benefits constraints on the shape of the Milky Way \citep{Ryan05,Ryan11,Pirzkal05,Pirzkal09,Holwerda14,van-Vledder16,Ryan17,Holwerda18}. The near-infrared observations of \emph{JWST} are sensitive to the lowest-mass objects (Y-dwarfs) throughout the width of the Galactic disk and halo. 

The randomized nature of the fields would allow for a first survey of Galactic disk and halo substellar objects, down to free-floating super-Jupiters \citep[e.g.,][]{Ryan17,Deacon18}.
These parallel searches for low/sub-stellar objects benefit from color information \citep{Holwerda18}, accurate astrometry, or grism information; but they derive their statistical power from the randomized fields. 

\subsection{Science Case C: Solar System Objects}

The existence of default parallels increases the chance to discover or follow-up on Solar System objects during deep stares. 
Meaningful, accidental science occurred during deep field observations with \textit{Spitzer}; for example, when \emph{Spitzer}/MIPS observations caught a passing asteroid \citep{Meadows04,Ryan15}.
The likelihood of near-infrared telescopes catching solar system objects is decidedly non-zero \citep{Kiss08}.
\emph{Spitzer} has performed several successful {\em targeted} observations of Solar System objects \citep{Stansberry04,Kelley13, Fernandez13a, Trilling16,Trilling17} as well as a study of solar system bodies in the ecliptic \citep{Meadows04}. As such, serendipitous and targeted programs on \emph{JWST} are fully complementary. 





Kuiper Belt objects (KBOs) can be identified in default parallels. High inclination KBOs, out of the ecliptic plane, are especially of interest \citep[e.g.,][]{Batygin16}. 
This would include objects similar to the KBO that was just visited by the New Horizons spacecraft, which has an R-band magnitude around 26-27, similar to the limiting magnitude for the default parallels program (Figure \ref{f:surveys}).
Moreover, \cite{Petit08} predict that $\sim1$ KBO object should exist per deg$^2$ with an $R$-band magnitude brighter than 23-23.5. \cite{Petit11} further estimated an observable population of KBOs with $m_g \sim 23-24.5$ per deg$^2$.
By covering a large number of deep, randomized fields, the default parallels program would have a significant possibility of discovering and quantifying the distribution of Solar System objects.

{KBO require at least two epochs for identification and we point out that many of the exoplanet observations are scheduled in multiple epochs as well. There will have to be some serendipity here (parallel on the same pointing and position angle). However, if the JWST default parallels represent a single epoch (e.g. F200W observations) the follow-up with EUCLID/WFIRST or HST could represent both the necessary secondary epoch as well as the H-band observation for H-band dropout of high redshift galaxies. }
\clearpage
\subsubsection{Small Bodies Discovery and Characterization}

As has been shown by the \emph{Kepler}, K2, and TESS surveys -- as well as prediction from \emph{JWST} yield estimates -- almost every field of view near the ecliptic is likely to be dominated by ``asteroid trails'' from main belt and outer solar system small bodies \citep{Wright10}. default parallels would provide a four-fold benefit to the planetary and astronomical communities:
\begin{enumerate}
	\item During deep imaging, the moving object trails would be used for discovery space. There are likely to be hundreds of thousands of faint solar system objects that have not yet been discovered; especially those emitting at the \emph{JWST}-MIRI wavelengths (15-24 $\mu$m).
    \item 
    Grism (WFSS) observations with the non-primary instruments to spectroscopically characterize moving objects in the solar system: KBOs, asteroids, and comets. 
    \item The randomized fields of view will provide direct capture of the distribution and population of small bodies in our solar system -- especially at high ecliptic latitudes; such as high-inclination, scattered objects \citep{Brown16a}.
    \item Because moving object trails are expected to be a significant source of astrophysical noise in the GTO/ERS/GO observations, having more detections of these sources would provide necessary information for mitigating these aberrations. 
\end{enumerate}

\subsection{Science Case D: Star-formation across cosmic time}

MIRI 24-$\mu$m observations to study star formation across cosmic time \citep{Brown17a,Clark18}, similar to \emph{Spitzer}/MIPS or \emph{WISE}/W4 but with sufficient resolution to resolve star-formation regions.
The 22-24 micron emission in a galaxy is strongly correlated to the total star-formation and can be used to accurately map local ($z<3$) star-formation rates \citep[see e.g.,][]{Cluver14,Cluver17}. Deep MIRI parallel observations will reveal where the star-formation is in the galaxy populations caught by the parallel observations. A similar consideration is being made for the SPICA science cases using this wavelength regime \citep[e.g.,][]{Bonato15,Gruppioni17}.

\begin{figure}
\begin{center}
	\includegraphics[width=\textwidth]{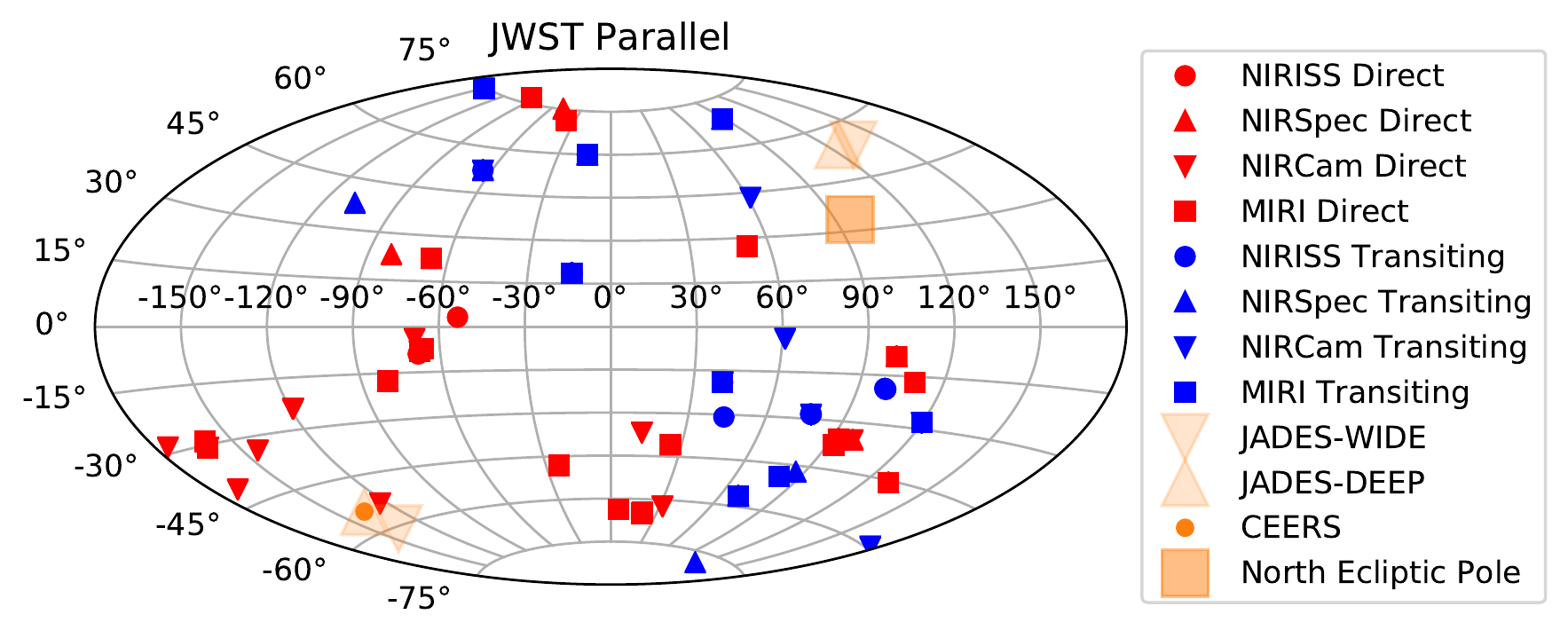}
	\caption{\label{f:instrument} The JWST exoplanet GTO programs projected on the sky in Galactic Coordinates with the {\em primary} instrument marked. }
\end{center}
\end{figure}

AGN information is encoded in this filter as well \citep{Jarrett11}. When MIRI is not the primary instrument for the exoplanet observations (see Figure \ref{f:instrument}), onboard data storage of MIRI default parallel imaging would not be a strong constraint. It could be the first field observations to supply star-formation and AGN studies with preliminary targets. The randomized nature of the pointings allow one to constrain the numbers of rare-and-bright sources.

\clearpage
\subsection{Science Case E: default parallel Time Series Observations}

Brown dwarf and exoplanet atmospheric characterization requires both high cadence (2-200 sec integrations) and large wavelength coverage (1-10 $\mu$m). 
To discover molecular abundances in colder, smaller atmospheres, predictions show that we need \emph{JWST} to be able to attain spectroscopic precisions on the order of 10-50ppm -- what is referred to as ``sub-50 ppm precision". 

Currently, it is unknown to what precision \emph{JWST} will be able to measure atmospheric features in the face of temporal, systematic noise sources; this is referred to as the ``noise floor" \citep{Greene16,Batalha18b}. If a bright star is known in the FOV of any non-primary instruments, while a time-series observation (TSO) is occurring on primary, and there are no immediate technical objections to conducting a second time series of a bright object ($K<15$), then default parallels could be used to characterize the star or possibly detect a transient source. This will have a range of long-term use from exoplanet studies and stellar characterization. The exoplanet community is attempting to attain precisions on their exoplanetary atmospheric signatures below 50ppm (possibly $<20$ppm) \citep{Greene16, Barstow16a,Barstow16,Barstow17, Batalha18a, Bean18}. 

With randomized fields of view to observe bright objects, a critical factor in predicting the efficacy of future TSO observations is how well each stellar type (FGKM), over several wavelengths, can attain sub-50 ppm precision. Both the commissioning and ERS programs are attempting to understand this yet unknown \emph{JWST} capability; but only a single target can be observed per program. 

Bright enough targets for TSO observations are not typically  close enough for both to fall into a second \emph{JWST} instrument field-of-view. 
However, even one bright TSO target observation in the proposed default parallels program will double the number of targets observed for long temporal baselines ($>10$k seconds).  
This critical information (sub-50 ppm efficacy) cannot be attained without successfully applying for a risky proposal (near saturation) in the usual GO program platform. 
Even one TSO target in a parallel would inform the sub-50 ppm efficacy for the entire exoplanet community. 

To date, several teams predict the need for detection of atmospheric features near 10ppm precision \citep[][and others]{Barstow16a,Barstow16,Barstow17,Kreidberg17, Batalha18b}.
With the wealth of possible planetary candidates from the \emph{TESS} satellite \citep{Ricker15}, the odds of two TSO observations near each other improve dramatically \citep{Sullivan15}.
This information will provide crucial input into in the planet formation and atmospheric predictions community as well.
A default parallel program with a Stellar Characterization/TSO component i.e. if a bright star does fall into a second instruments view the program would switch to TSO, would greatly inform the community to what level of precision \emph{JWST} predictions can reach.

Depending on the brightness of the object ($m_K\sim5,8,10,12$), if imaging is expected to saturate, default parallels could perform grism TSO or WFSS observation instead of imaging for both NIRISS and NIRCam. This would minimize risk of saturation and maximize potential for stellar characterization at longer wavelengths;  building up a template selection for \emph{JWST} for all users.
%
%
%

An equally necessary benefit for default parallel TSO observations is that the greatest source of uncertainty with high precision transiting exoplanet observations is that the stellar spectra are not as well characterized as is necessary to produce exoplanetary absolute abundance measurements.
Stellar characterization through grism default parallels would benefit stellar, brown dwarf, and exoplanet observations by developing a database of stellar templates for the entire JWST program to use.

\subsection{Science Case F: Grism Observations}

As an alternative to the imaging options presented above, grism observations for the same science cases (Galactic brown dwarf population, high-redshift galaxy populations) could be considered. 
Deep grism stares would identify brown dwarfs \citep{Holwerda14,Ryan17,Deacon18} as well as $z\sim11$ galaxies \citep{Oesch16}, of which there is a remarkable dearth \citep{Oesch18a}. The WFC3 Infrared Spectroscopic Parallel Survey (WISP) default parallel grism survey with \emph{HST}/WFC3 \citep{Atek10} has successfully characterized the intermediate redshift population \citep[e.g.,][]{Atek11,Atek14,Malkan13,Bedregal13}, predicted more use of grism spectroscopy in NIR surveys \citep{Colbert13} and identified some Milky Way Halo objects \citep{Masters12a}. Grism observations also hold the potential to cleanly separate AGN and star-formation contributions in intermediate redshift galaxy populations \citep{Trump11, Bridge16}. 

All four of the instruments on \textit{JWST} employ slitless spectroscopic modes. In particular, NIRCam ($R\sim1600$,  $2.4<\lambda<5$ $\mu$m) and NIRISS ($R\sim150$, $0.8<\lambda<2.2$ $\mu$m) grisms both have a wide-field mode similar to the \emph{HST} WFC3 and ACS grisms. Leveraging parallel \emph{JWST} observations, as has been done with \emph{HST} grism surveys, will be an important undertaking to maximize scientific results. For example, doubly-ionized oxygen has recently been detected at $z\sim9$ using ALMA \citep{Hashimoto18a}
, and the ability to spatially resolve [O~II] out to $z\sim12$ may revolutionize our understanding of star formation in the earliest galaxies, less than 250 million years after the Big Bang. We will also be able to probe the [O III]/H$\beta$ ratio out to $z\sim9$, facilitating testing of theories about black hole seeds in the early Universe and whether the existence of low-luminosity AGN, which have been found in significant numbers at $4<z<6.5$ \citep{Giallongo15a} played a significant role in reionization.

The longer exposures without many intermezzos for default parallels would lend themselves preferentially for grism observations.

\section{Exposure Times \& Survey Area}


We consider default parallel imaging as the simplest science case, single band observations. A TSO default parallel is likely rare and grism observations with single orientations are more difficult to simply predict the scientific yield, but should be similar to a multi-band survey.

\begin{figure}
\begin{center}
	\includegraphics[width=\textwidth]{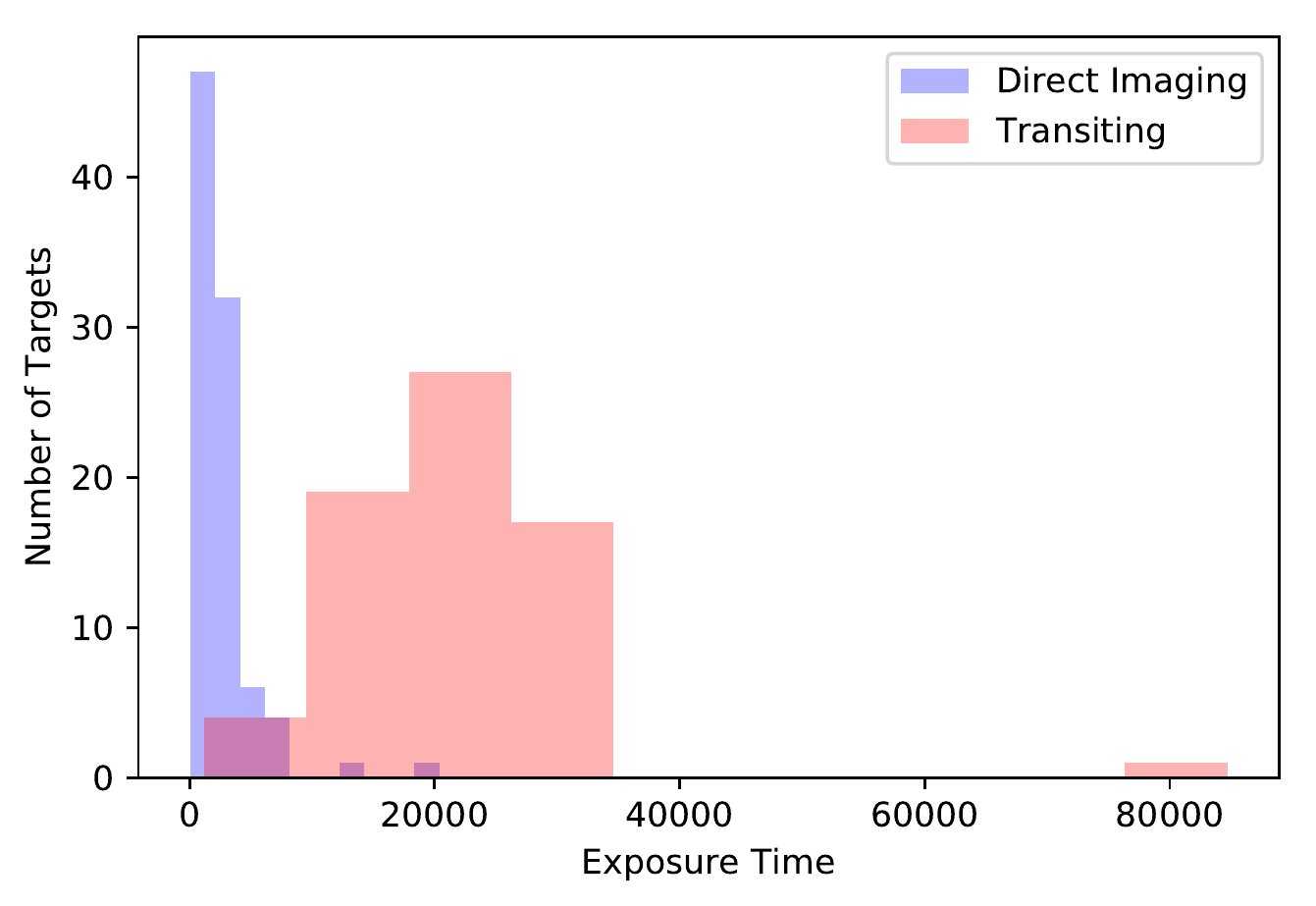}
	\caption{\label{f:histtime} The exposure time histogram of the direct imaging and transiting JWST GTO programs. }
\end{center}
\end{figure}

The mean exposure time for the GTO direct imaging is 2.5k 
seconds and for the GTO transiting programs is 22k 
seconds. Figure \ref{f:histtime} shows the distribution of exposure times asked for in the GTO programs. One expects GO observations to span lower integration times. The transiting programs offer the best options for a series of deep fields, possibly using multiple filters (if the transit observations include scheduled breaks, long enough to allow for filter changes). 

%
%
%
%
A couple of thousand seconds of exposure time with NIRCam or NIRISS can reach depths of $m\sim28.0$ in F200W/F277W 
and $M_{UV} \sim -19.5$ galaxy at $9<z<13$ \citep[cf. CEERS proposals, PI Finkelstein][]{CEERS}. {The JWST default parallel observations provide the redward band for H-band dropout selections using HST, WFIRST or EUCLID auxiliary/follow-up data.}

These default parallel observations equivalent to the GTO's medium depth imaging surveys. Default parallels with the GTO transit observation programs could rival dedicated medium-deep fields, and over randomized fields of view.

\begin{figure*}
\begin{center}
	\includegraphics[width=\textwidth]{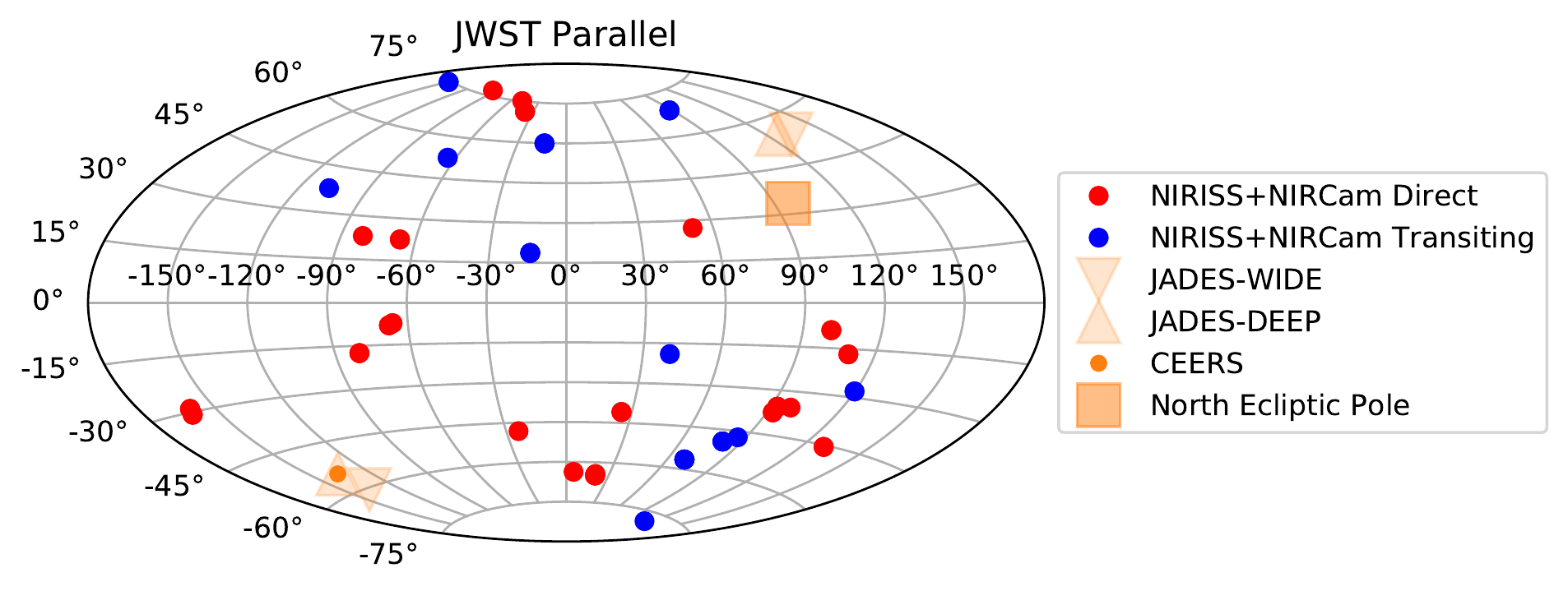}
	\caption{\label{f:twocam} The JWST exoplanet GTO programs projected on the sky in Galactic Coordinates with the availability of both NIRISS and NIRCAM as {\em non-primary} instrument marked. }
\end{center}
\end{figure*}

There are totals of 17k and 10k seconds available with both NIRISS and NIRCam as the parallel instruments over 19 fields (Figure \ref{f:twocam}). Only one field looks to have too low of a Galactic latitude to be of extra-Galactic use and four may suffer from stellar crowding ($b<15^\circ$). 

\begin{figure}
\begin{center}
	\includegraphics[width=\textwidth]{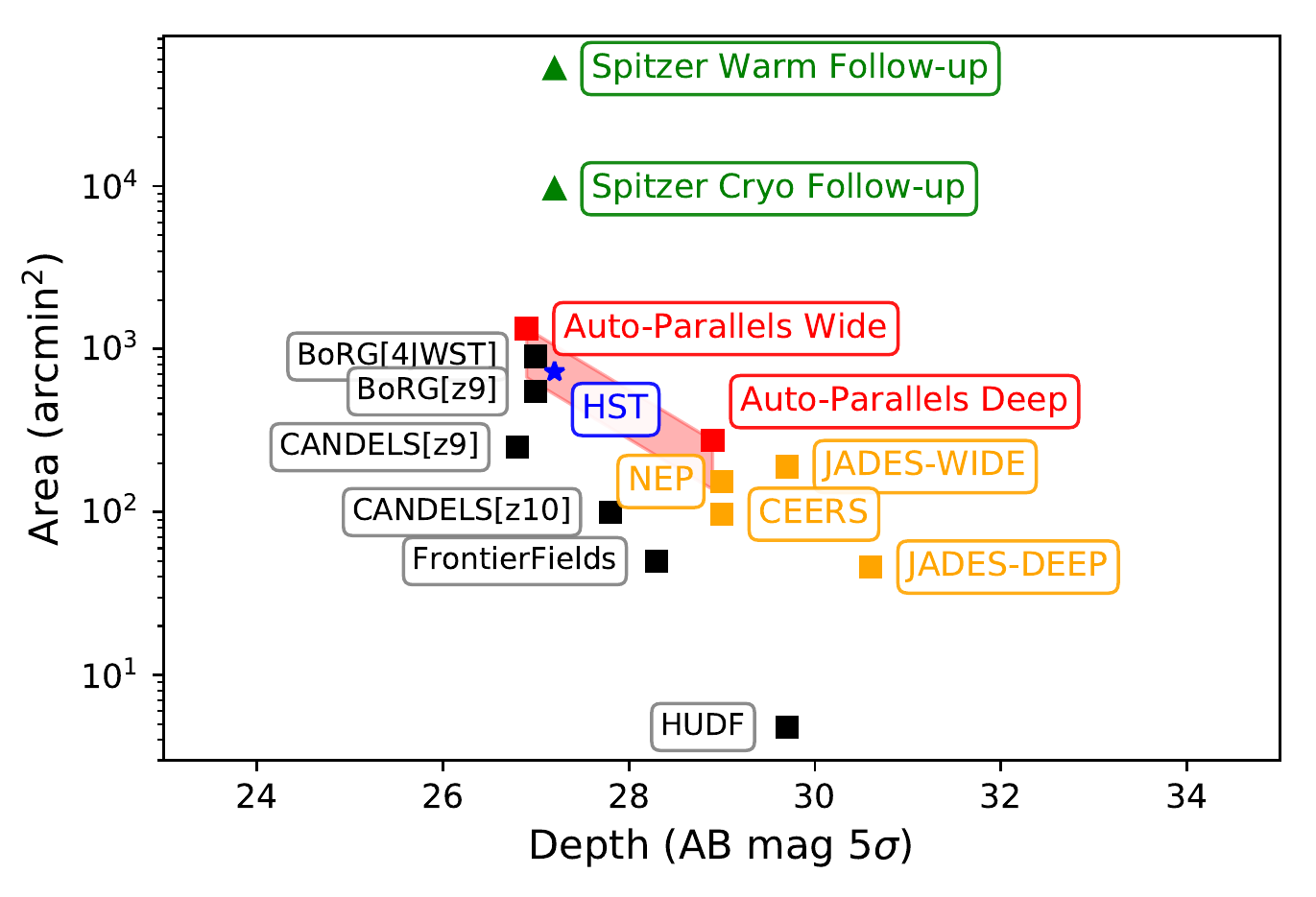}
	\caption{\label{f:surveys} The approximate limiting depth and area covered by the existing \emph{HST} surveys (black), the GTO CEERS survey \citep[P.I. S. Finkelstein][]{CEERS}, JADES \citep[P.I. M. Rieke][]{Williams18,Rieke19}, and NEP \citep[P.I. R. Windhorst][]{Jansen18} and the possible depth and coverage of default parallels using just NIRCam and NIRISS combined pointings. The Deep and Wide scenarios are ones where the principal observation's roll angle is kept constant (deep) or varied significantly (wide). The shaded area is the range of expected depths and area covered for default parallels. These are the approximate depths and coverage for a single filter. GO observations could potentially widen the default parallel Wide option by a factor two or more.
    Assuming all \emph{Spitzer} Cryo and Warm mission targets are followed up with instruments other than NIRCam and NIRISS in GO programs, a shallow tier to the parallel surveys could be added. 
    }
\end{center}
\end{figure}

This default parallel opportunity alone would be between 275 arcmin$^2$ to almost a {\em half a square degree} survey, depending on the primary observation orientation angle constraints and the number of approved GO proposals. If the orientation angle is kept fixed, then the resulting parallel observation will be much deeper; if there is significant variation, the parallels will provide a wider survey (Figure \ref{f:surveys}).

Each of the HST deep surveys of the high-redshift Universe (CANDELS, Frontier Fields, etc) are less than 400 arcmin$^2$ each. 
Three GTO programs are the first deep, multi-wavelength program with JWST to showcase all the instruments capabilities: the Cosmic Evolution Early Release Science \citep[CEERS, P.I. S. Finkelstein][]{CEERS}, the JWST Advanced Deep Extragalactic Survey \citep[JADES][]{Williams18,Rieke19}, and the North Ecliptic Pole \citep[NEP][]{Jansen18} GTO deep fields. 

In a single filter, default parallels would expand the discovery space significantly (Figure \ref{f:surveys}). The value of these observations will be in both comparing the known source fields to default parallels' randomized fields of view, and plausibly discovering new deep targets to follow up with regular GO opportunities.


\section{Discussion}
Parallel mode observed random pointings reduce cosmic variance more effectively than single continuous fields of a typical wedding cake observing strategy \citep{Trenti08}. Hence, we argue that the opportunity for an default parallels survey will explore an unique and rich discovery space of high redshift, TSOs, Solar System objects, and Galactic structure.

Random, parallel fields can counter two issues that confront high-redshift searches: cosmic variance and human bias in deep field selection. Cosmic variance remains a dominant source of uncertainty in the relatively small areas surveyed by instruments like \emph{HST} and \emph{JWST} \citep{Driver10}. To counter that, larger continuous areas ($>1^\circ$) can be considered but these remain observationally expensive. A randomized sampling does an equally or better job of countering cosmic variance \citep{Trenti08}. 
The current deep observations by \emph{HST}, e.g. CANDELS \citep{Koekemoer11,Grogin11}, are all focused on those fields for which much needed ancillary data is already available. This perpetuates a bias for low Galactic cirrus regions identified decades ago. In essence, the choice for \emph{JWST} fields was locked in at that time. Deep, randomly chosen, fields can act as a control on deep fields dictated by legacy.
%
%
%
Our nominal case is one that uses only the F200W or the F277W filters to perform a uniform default parallels survey. However, if there are no issues with roll angle, a series of filters can be adopted (e.g., F150W, F200W, F277W and F356W) to combine into a photometric redshift search for high-redshift galaxies. If, for example, a transiting observation is done with intermissions, then that time can be used to switch filters in the non-primary instruments.
{The principle idea for default parallels is to be a program that is executed with only a few simple rules dictating its observation strategy for ease of implementation.} 



\begin{figure}
\begin{center}
	\includegraphics[width=\textwidth]{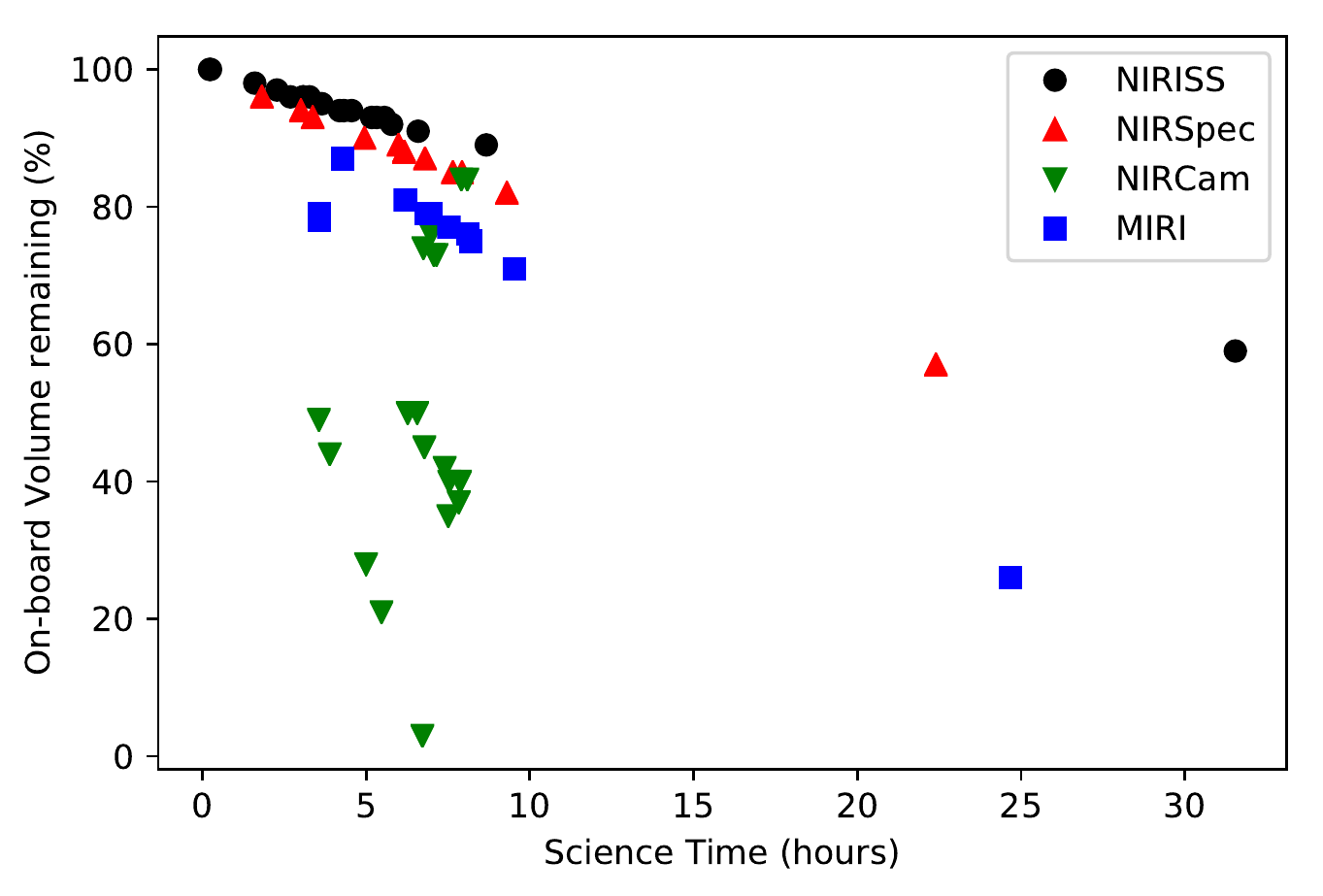}
	\caption{\label{f:memory} The remaining onboard memory for each ERS program  with the primary instrument labelled. A fraction of the ERS programs use more than half the onboard memory, predominantly NIRCam programs, but two of the three very long duration ERS observations leave 50\% of the memory for potential parallel use. 
    }
\end{center}
\end{figure}

There are several objections to an default parallel program that is only a single design. The program would use resources for \emph{JWST}, both human and spacecraft (onboard storage and power), as well as MAST archive storage and support. Some of these are trivial but some are not and need a solid reason for their commitment. A single program would cost less but also is less flexible in its science case. It would minimize parallel use from the pool of available time for which JWST users could propose. {We argue here that at present these are spacecraft hours that are not being considered since parallel observations are not allowed in concert with any non-imaging observation. }

These deep exposures are also default parallel opportunities for dark exposure time. These are critical measurements of the detectors themselves and time or parallel mode time will need to be scheduled for separately. This need not be mutually exclusive: both a dark and a parallel observations could potentially be done during the same primary program. Only the onboard data storage limits this dual use. {Figure \ref{f:memory} shows the science time and remaining onboard memory for the ERS programs. Most use less than 50\% of the storage capability of JWST, even for long exposures ($>20$ hours) with no primary instrument excluded.\footnote{The NIRCam observations are the most memory-intensive but the default parallels with NIRCam need not be: these are likely only a few exposures, to rid of cosmic rays.}
The remaining free space allows in principle for default parallel observations, dark frames, or both for the remaining instruments. We list the primary instrument, science and charged time, and remaining storage in Table \ref{t:ers:usage} for the ERS exoplanet programs.  }

These observations would not be of the same quality as dedicated observations: dithering strategy --if any-- will be dictated by the primary science instrument. Integration times (to limit buffer usage) could be modified to minimize data storage (i.e.~DEEP8 observations). The filter choice must be fixed, because filter changes are not possible during a time-series observations, due to the incurred associated dither (i.e.~vibrations from filter wheel motions). In the case of long period transiting campaigns (i.e.~phase curve observations) and high contrast imaging with multiple roll angles, then multiple filter information might be obtained in the parallel instrument as well. 
In the case of grism default parallel observations (if these are added to the program), roll angles will be constrained by the primary observation and no second angle will necessarily be available. 

These data will not take the place of dedicated high-redshift observation campaigns that are on the docket for fields with a wealth of complementary (existing) data.
Examples of similar programs with \emph{HST} (ACS snapshot and the WFPC2 B-band pure parallel survey) build a legacy archive that often only reveal their use much later. 

For example, the original ACS "schedule-gap" SNAP program \citep[14840, PI: Bellini) ][]{Bellini16b,Bellini16a,Bellini17} fortuitously monitored the NGC 4993 progenitor of the 2017 LIGO kilonova, which constrained the pre-nova mass \citep{Cowperthwaite17,Levan17,Pan17}. 
Similarly, a dedicated WFPC2 default parallel program produced long-term value for the archive \citep{Casertano02,Wadadekar06}. Such long-term value has shown to be a critical component of the Hubble Space Telescope's legacy and scientific success \citep[e.g.][]{Peek17}.
%
%

The current, accepted \emph{JWST} Parallel program does not permit parallel observations during transiting exoplanet observations or with multiple instruments. Individually requested parallels would require each team to decide the observation strategy per parallel field. default parallels would use an objective, community 
designed decision tree to dictate the maximum scientific efficiency per default parallel field of view, during an exoplanet primary. 

We argue here that the long exposure times necessary for exoplanet observations are an opportunity that cannot be left fallow. Either through an automated observing program --the proposed default parallels or as part of a vigorous and supported proposal process, the parallel time during exoplanet primary observations is a rich resource that needs to be exploited. 

Although direct and transiting exoplanet observations are highly likely to occur throughout the lifetime of \textit{JWST}, if we wait to enact this default parallels program after Cycle-1, we would lose access to parallel observations for the majority of GTO/ERS observations; the only currently existing, approved programs. This would then push back any future follow-up and characterization of plausible detections in our default parallel fields; limiting the value of each FOV.  The sooner \textit{JWST} project enacts a default parallel program, the greater the legacy of these fortuitous observations will be. 
%
%
%

We propose default parallels as a low-cost program to deliver a homogeneous and practical data-set to serve a wide variety of science cases, building on the statistical strength of random sampling.

\section*{Acknowledgements}
 
We thank Karl Gordon for the initial discussions on this subject and the suggestion to write this paper.
%
%
%
%
%
This research made use of Astropy, a community-developed core Python package for Astronomy \citep{Astropy-Collaboration13a}. This research made use of matplotlib, a Python library for publication quality graphics \citep{Hunter07}. PyRAF is a product of the Space Telescope Science Institute, which is operated by AURA for NASA. This research made use of both SciPy \citep{scipy}.


\begin{thebibliography}{82}
\expandafter\ifx\csname natexlab\endcsname\relax\def\natexlab#1{#1}\fi

\bibitem[{{Astropy Collaboration} {et~al.}(2013){Astropy Collaboration},
  {Robitaille}, {Tollerud}, {Greenfield}, {Droettboom}, {Bray}, {Aldcroft},
  {Davis}, {Ginsburg}, {Price-Whelan}, {Kerzendorf}, {Conley}, {Crighton},
  {Barbary}, {Muna}, {Ferguson}, {Grollier}, {Parikh}, {Nair}, {Unther},
  {Deil}, {Woillez}, {Conseil}, {Kramer}, {Turner}, {Singer}, {Fox}, {Weaver},
  {Zabalza}, {Edwards}, {Azalee Bostroem}, {Burke}, {Casey}, {Crawford},
  {Dencheva}, {Ely}, {Jenness}, {Labrie}, {Lim}, {Pierfederici}, {Pontzen},
  {Ptak}, {Refsdal}, {Servillat}, \& {Streicher}}]{Astropy-Collaboration13a}
{Astropy Collaboration} {et~al.} 2013, \aap, 558, A33

\bibitem[{{Atek} {et~al.}(2014){Atek}, {Kneib}, {Pacifici}, {Malkan},
  {Charlot}, {Lee}, {Bedregal}, {Bunker}, {Colbert}, {Dressler}, {Hathi},
  {Lehnert}, {Martin}, {McCarthy}, {Rafelski}, {Ross}, {Siana}, \&
  {Teplitz}}]{Atek14}
{Atek}, H. {et~al.} 2014, \apj, 789, 96

\bibitem[{{Atek} {et~al.}(2010){Atek}, {Malkan}, {McCarthy}, {Teplitz},
  {Scarlata}, {Siana}, {Henry}, {Colbert}, {Ross}, {Bridge}, {Bunker},
  {Dressler}, {Fosbury}, {Martin}, \& {Shim}}]{Atek10}
---. 2010, \apj, 723, 104

\bibitem[{{Atek} {et~al.}(2011){Atek}, {Siana}, {Scarlata}, {Malkan},
  {McCarthy}, {Teplitz}, {Henry}, {Colbert}, {Bridge}, {Bunker}, {Dressler},
  {Fosbury}, {Hathi}, {Martin}, {Ross}, \& {Shim}}]{Atek11}
---. 2011, ArXiv e-prints

\bibitem[{{Barstow} {et~al.}(2017){Barstow}, {Aigrain}, {Irwin}, \&
  {Sing}}]{Barstow17}
{Barstow}, J.~K. {et~al.} 2017, \apj, 834, 50

\bibitem[{{Barstow} \& {Irwin}(2016)}]{Barstow16}
{Barstow}, J.~K., \& {Irwin}, P.~G.~J. 2016, \mnras, 461, L92

\bibitem[{{Barstow} {et~al.}(2016){Barstow}, {Irwin}, {Kendrew}, \&
  {Aigrain}}]{Barstow16a}
{Barstow}, J.~K. {et~al.} 2016, in Space Telescopes and Instrumentation 2016:
  Optical, Infrared, and Millimeter Wave, Vol. 9904, 99043P

\bibitem[{{Batalha} {et~al.}(2018{\natexlab{a}}){Batalha}, {Lewis}, {Line},
  {Valenti}, \& {Stevenson}}]{Batalha18b}
{Batalha}, N.~E. {et~al.} 2018{\natexlab{a}}, \apj, 856, L34

\bibitem[{{Batalha} {et~al.}(2018{\natexlab{b}}){Batalha}, {Smith}, {Lewis},
  {Marley}, {Fortney}, \& {Macintosh}}]{Batalha18a}
---. 2018{\natexlab{b}}, \aj, 156, 158

\bibitem[{{Batygin} \& {Brown}(2016)}]{Batygin16}
{Batygin}, K., \& {Brown}, M.~E. 2016, \apjl, 833, L3

\bibitem[{{Bean} {et~al.}(2018){Bean}, {Stevenson}, {Batalha},
  {Berta-Thompson}, {Kreidberg}, {Crouzet}, {Benneke}, {Line}, {Sing},
  {Wakeford}, {Knutson}, {Kempton}, {D{\'e}sert}, {Crossfield}, {Batalha}, {de
  Wit}, {Parmentier}, {Harrington}, {Moses}, {Lopez-Morales}, {Alam}, {Blecic},
  {Bruno}, {Carter}, {Chapman}, {Decin}, {Dragomir}, {Evans}, {Fortney},
  {Fraine}, {Gao}, {Garc{\'\i}a Mu{\~n}oz}, {Gibson}, {Goyal}, {Heng}, {Hu},
  {Kendrew}, {Kilpatrick}, {Krick}, {Lagage}, {Lendl}, {Louden}, {Madhusudhan},
  {Mandell}, {Mansfield}, {May}, {Morello}, {Morley}, {Nikolov}, {Redfield},
  {Roberts}, {Schlawin}, {Spake}, {Todorov}, {Tsiaras}, {Venot}, {Waalkes},
  {Wheatley}, {Zellem}, {Angerhausen}, {Barrado}, {Carone}, {Casewell},
  {Cubillos}, {Damiano}, {de Val-Borro}, {Drummond}, {Edwards}, {Endl},
  {Espinoza}, {France}, {Gizis}, {Greene}, {Henning}, {Hong}, {Ingalls}, {Iro},
  {Irwin}, {Kataria}, {Lahuis}, {Leconte}, {Lillo-Box}, {Lines}, {Lothringer},
  {Mancini}, {Marchis}, {Mayne}, {Palle}, {Rauscher}, {Roudier}, {Shkolnik},
  {Southworth}, {Swain}, {Taylor}, {Teske}, {Tinetti}, {Tremblin}, {Tucker},
  {van Boekel}, {Waldmann}, {Weaver}, \& {Zingales}}]{Bean18}
{Bean}, J.~L. {et~al.} 2018, Publications of the Astronomical Society of the
  Pacific, 130, 114402

\bibitem[{{Bedregal} {et~al.}(2013){Bedregal}, {Scarlata}, {Henry}, {Atek},
  {Rafelski}, {Teplitz}, {Dominguez}, {Siana}, {Colbert}, {Malkan}, {Ross},
  {Martin}, {Dressler}, {Bridge}, {Hathi}, {Masters}, {McCarthy}, \&
  {Rutkowski}}]{Bedregal13}
{Bedregal}, A.~G. {et~al.} 2013, \apj, 778, 126

\bibitem[{{Bellini}(2016{\natexlab{a}})}]{Bellini16b}
{Bellini}, A. 2016{\natexlab{a}}, {Extended F814W Schedule Gap Pilot}, HST
  Proposal

\bibitem[{{Bellini}(2016{\natexlab{b}})}]{Bellini16a}
---. 2016{\natexlab{b}}, {Schedule Gap Pilot}, HST Proposal

\bibitem[{{Bellini} {et~al.}(2017){Bellini}, {Grogin}, {Hathi}, \&
  {Brown}}]{Bellini17}
{Bellini}, A. {et~al.} 2017, Instrument Science Report

\bibitem[{{Bernard} {et~al.}(2016){Bernard}, {Carrasco}, {Trenti}, {Oesch},
  {Wu}, {Bradley}, {Schmidt}, {Bouwens}, {Calvi}, {Mason}, {Stiavelli}, \&
  {Treu}}]{Bernard16}
{Bernard}, S.~R. {et~al.} 2016, \apj, 827, 76

\bibitem[{{Bonato}(2015)}]{Bonato15}
{Bonato}, M. 2015, PhD thesis, University of Padua

\bibitem[{{Bradley} {et~al.}(2012){Bradley}, {Trenti}, {Oesch}, {Stiavelli},
  {Treu}, {Bouwens}, {Shull}, {Holwerda}, \& {Pirzkal}}]{Bradley12}
{Bradley}, L.~D. {et~al.} 2012, \apj, 760, 108

\bibitem[{{Bridge} {et~al.}(2016){Bridge}, {Zeimann}, {Trump}, {Gronwall},
  {Ciardullo}, {Fox}, \& {Schneider}}]{Bridge16}
{Bridge}, J.~S. {et~al.} 2016, \apj, 826, 172

\bibitem[{{Brown} \& {Batygin}(2016)}]{Brown16a}
{Brown}, M.~E., \& {Batygin}, K. 2016, \apjl, 824, L23

\bibitem[{{Brown} {et~al.}(2017){Brown}, {Moustakas}, {Kennicutt}, {Bonne},
  {Intema}, {de Gasperin}, {Boquien}, {Jarrett}, {Cluver}, {Smith}, {da Cunha},
  {Imanishi}, {Armus}, {Brandl}, \& {Peek}}]{Brown17a}
{Brown}, M.~J.~I. {et~al.} 2017, \apj, 847, 136

\bibitem[{{Calvi} {et~al.}(2016){Calvi}, {Trenti}, {Stiavelli}, {Oesch},
  {Bradley}, {Schmidt}, {Coe}, {Brammer}, {Bernard}, {Bouwens}, {Carrasco},
  {Carollo}, {Holwerda}, {MacKenty}, {Mason}, {Shull}, \& {Treu}}]{Calvi16}
{Calvi}, V. {et~al.} 2016, \apj, 817, 120

\bibitem[{{Cameron} {et~al.}(2019){Cameron}, {Trenti}, {Livermore}, \& {van der
  Velden}}]{Cameron19}
{Cameron}, A.~J. {et~al.} 2019, \mnras, 483, 1922

\bibitem[{{Casertano}(2002)}]{Casertano02}
{Casertano}, S. 2002, {The WFPC2 Archival Parallels}, HST Proposal

\bibitem[{{Clark} {et~al.}(2018){Clark}, {Verstocken}, {Bianchi}, {Fritz},
  {Viaene}, {Smith}, {Baes}, {Casasola}, {Cassara}, {Davies}, {De Looze}, {De
  Vis}, {Evans}, {Galametz}, {Jones}, {Lianou}, {Madden}, {Mosenkov}, \&
  {Xilouris}}]{Clark18}
{Clark}, C.~J.~R. {et~al.} 2018, \aap, 609, A37

\bibitem[{{Cluver} {et~al.}(2017){Cluver}, {Jarrett}, {Dale}, {Smith},
  {August}, \& {Brown}}]{Cluver17}
{Cluver}, M.~E. {et~al.} 2017, \apj, 850, 68

\bibitem[{{Cluver} {et~al.}(2014){Cluver}, {Jarrett}, {Hopkins}, {Driver},
  {Liske}, {Gunawardhana}, {Taylor}, {Robotham}, {Alpaslan}, {Baldry}, {Brown},
  {Peacock}, {Popescu}, {Tuffs}, {Bauer}, {Bland-Hawthorn}, {Colless},
  {Holwerda}, {Lara-L{\'o}pez}, {Leschinski}, {L{\'o}pez-S{\'a}nchez},
  {Norberg}, {Owers}, {Wang}, \& {Wilkins}}]{Cluver14}
---. 2014, \apj, 782, 90

\bibitem[{{Colbert} {et~al.}(2013){Colbert}, {Teplitz}, {Atek}, {Bunker},
  {Rafelski}, {Ross}, {Scarlata}, {Bedregal}, {Dominguez}, {Dressler}, {Henry},
  {Malkan}, {Martin}, {Masters}, {McCarthy}, \& {Siana}}]{Colbert13}
{Colbert}, J.~W. {et~al.} 2013, \apj, 779, 34

\bibitem[{{Cowperthwaite} {et~al.}(2017){Cowperthwaite}, {Berger}, {Villar},
  {Metzger}, {Nicholl}, {Chornock}, {Blanchard}, {Fong}, {Margutti},
  {Soares-Santos}, {Alexander}, {Allam}, {Annis}, {Brout}, {Brown}, {Butler},
  {Chen}, {Diehl}, {Doctor}, {Drout}, {Eftekhari}, {Farr}, {Finley}, {Foley},
  {Frieman}, {Fryer}, {Garc{\'{\i}}a-Bellido}, {Gill}, {Guillochon}, {Herner},
  {Holz}, {Kasen}, {Kessler}, {Marriner}, {Matheson}, {Neilsen}, {Quataert},
  {Palmese}, {Rest}, {Sako}, {Scolnic}, {Smith}, {Tucker}, {Williams},
  {Balbinot}, {Carlin}, {Cook}, {Durret}, {Li}, {Lopes}, {Louren{\c c}o},
  {Marshall}, {Medina}, {Muir}, {Mu{\~n}oz}, {Sauseda}, {Schlegel}, {Secco},
  {Vivas}, {Wester}, {Zenteno}, {Zhang}, {Abbott}, {Banerji}, {Bechtol},
  {Benoit-L{\'e}vy}, {Bertin}, {Buckley-Geer}, {Burke}, {Capozzi}, {Carnero
  Rosell}, {Carrasco Kind}, {Castander}, {Crocce}, {Cunha}, {D'Andrea}, {da
  Costa}, {Davis}, {DePoy}, {Desai}, {Dietrich}, {Drlica-Wagner}, {Eifler},
  {Evrard}, {Fernandez}, {Flaugher}, {Fosalba}, {Gaztanaga}, {Gerdes},
  {Giannantonio}, {Goldstein}, {Gruen}, {Gruendl}, {Gutierrez}, {Honscheid},
  {Jain}, {James}, {Jeltema}, {Johnson}, {Johnson}, {Kent}, {Krause}, {Kron},
  {Kuehn}, {Nuropatkin}, {Lahav}, {Lima}, {Lin}, {Maia}, {March}, {Martini},
  {McMahon}, {Menanteau}, {Miller}, {Miquel}, {Mohr}, {Neilsen}, {Nichol},
  {Ogando}, {Plazas}, {Roe}, {Romer}, {Roodman}, {Rykoff}, {Sanchez},
  {Scarpine}, {Schindler}, {Schubnell}, {Sevilla-Noarbe}, {Smith}, {Smith},
  {Sobreira}, {Suchyta}, {Swanson}, {Tarle}, {Thomas}, {Thomas}, {Troxel},
  {Vikram}, {Walker}, {Wechsler}, {Weller}, {Yanny}, \&
  {Zuntz}}]{Cowperthwaite17}
{Cowperthwaite}, P.~S. {et~al.} 2017, \apjl, 848, L17

\bibitem[{{Deacon}(2018)}]{Deacon18}
{Deacon}, N.~R. 2018, \mnras, 481, 447

\bibitem[{{Driver} \& {Robotham}(2010)}]{Driver10}
{Driver}, S.~P., \& {Robotham}, A.~S.~G. 2010, \mnras, 407, 2131

\bibitem[{{Fern{\'a}ndez} {et~al.}(2013){Fern{\'a}ndez}, {Kelley}, {Lamy},
  {Toth}, {Groussin}, {Lisse}, {A'Hearn}, {Bauer}, {Campins}, {Fitzsimmons},
  {Licandro}, {Lowry}, {Meech}, {Pittichov{\'a}}, {Reach}, {Snodgrass}, \&
  {Weaver}}]{Fernandez13a}
{Fern{\'a}ndez}, Y.~R. {et~al.} 2013, \icarus, 226, 1138

\bibitem[{{Finkelstein} {et~al.}(2017){Finkelstein}, {Dickinson}, {Ferguson},
  {Grazian}, {Grogin}, {Kartaltepe}, {Kewley}, {Kocevski}, {Koekemoer}, \&
  {Lotz}}]{CEERS}
{Finkelstein}, S. {et~al.} 2017, {The Cosmic Evolution Early Release Science
  (CEERS) Survey}, JWST Proposal ID 1345. Cycle 0 Early Release Scienc

\bibitem[{{Giallongo} {et~al.}(2015){Giallongo}, {Grazian}, {Fiore}, {Fontana},
  {Pentericci}, {Vanzella}, {Dickinson}, {Kocevski}, {Castellano}, {Cristiani},
  {Ferguson}, {Finkelstein}, {Grogin}, {Hathi}, {Koekemoer}, {Newman}, \&
  {Salvato}}]{Giallongo15a}
{Giallongo}, E. {et~al.} 2015, \aap, 578, A83

\bibitem[{{Greene} {et~al.}(2016){Greene}, {Line}, {Montero}, {Fortney},
  {Lustig-Yaeger}, \& {Luther}}]{Greene16}
{Greene}, T.~P. {et~al.} 2016, \apj, 817, 17

\bibitem[{{Grogin} {et~al.}(2011){Grogin}, {Kocevski}, {Faber}, {Ferguson},
  {Koekemoer}, {Riess}, {Acquaviva}, {Alexander}, {Almaini}, {Ashby}, {Barden},
  {Bell}, {Bournaud}, {Brown}, {Caputi}, {Casertano}, {Cassata}, {Castellano},
  {Challis}, {Chary}, {Cheung}, {Cirasuolo}, {Conselice}, {Roshan Cooray},
  {Croton}, {Daddi}, {Dahlen}, {Dav{\'e}}, {de Mello}, {Dekel}, {Dickinson},
  {Dolch}, {Donley}, {Dunlop}, {Dutton}, {Elbaz}, {Fazio}, {Filippenko},
  {Finkelstein}, {Fontana}, {Gardner}, {Garnavich}, {Gawiser}, {Giavalisco},
  {Grazian}, {Guo}, {Hathi}, {H{\"a}ussler}, {Hopkins}, {Huang}, {Huang},
  {Jha}, {Kartaltepe}, {Kirshner}, {Koo}, {Lai}, {Lee}, {Li}, {Lotz}, {Lucas},
  {Madau}, {McCarthy}, {McGrath}, {McIntosh}, {McLure}, {Mobasher},
  {Moustakas}, {Mozena}, {Nandra}, {Newman}, {Niemi}, {Noeske}, {Papovich},
  {Pentericci}, {Pope}, {Primack}, {Rajan}, {Ravindranath}, {Reddy}, {Renzini},
  {Rix}, {Robaina}, {Rodney}, {Rosario}, {Rosati}, {Salimbeni}, {Scarlata},
  {Siana}, {Simard}, {Smidt}, {Somerville}, {Spinrad}, {Straughn}, {Strolger},
  {Telford}, {Teplitz}, {Trump}, {van der Wel}, {Villforth}, {Wechsler},
  {Weiner}, {Wiklind}, {Wild}, {Wilson}, {Wuyts}, {Yan}, \& {Yun}}]{Grogin11}
{Grogin}, N.~A. {et~al.} 2011, \apjs, 197, 35

\bibitem[{{Gruppioni} {et~al.}(2017){Gruppioni}, {Ciesla}, {Hatziminaoglou},
  {Pozzi}, {Rodighiero}, {Santini}, {Armus}, {Baes}, {Braine}, {Charmandaris},
  {Clements}, {Christopher}, {Dannerbauer}, {Efstathiou}, {Egami},
  {Fern{\'a}ndez-Ontiveros}, {Fontanot}, {Franceschini},
  {Gonz{\'a}lez-Alfonso}, {Griffin}, {Kaneda}, {Marchetti}, {Monaco},
  {Nakagawa}, {Onaka}, {Papadopoulos}, {Pearson}, {P{\'e}rez-Fournon},
  {Per{\'e}z-Gonz{\'a}lez}, {Roelfsema}, {Scott}, {Serjeant}, {Spinoglio},
  {Vaccari}, {van der Tak}, {Vignali}, {Wang}, \& {Wada}}]{Gruppioni17}
{Gruppioni}, C. {et~al.} 2017, \pasa, 34, e055

\bibitem[{{Hashimoto} {et~al.}(2018){Hashimoto}, {Laporte}, {Mawatari},
  {Ellis}, {Inoue}, {Zackrisson}, {Roberts-Borsani}, {Zheng}, {Tamura},
  {Bauer}, {Fletcher}, {Harikane}, {Hatsukade}, {Hayatsu}, {Matsuda}, {Matsuo},
  {Okamoto}, {Ouchi}, {Pell{\'o}}, {Rydberg}, {Shimizu}, {Taniguchi},
  {Umehata}, \& {Yoshida}}]{Hashimoto18a}
{Hashimoto}, T. {et~al.} 2018, \nat, 557, 392

\bibitem[{{Holwerda} {et~al.}(2018){Holwerda}, {Bridge}, {Ryan}, {Kenworthy},
  {Pirzkal}, {Andersen}, {Wilkins}, {Smit}, {Bernard}, {Meshkat}, {Steele}, \&
  {Bouwens}}]{Holwerda18}
{Holwerda}, B.~W. {et~al.} 2018, \aap, 620, A132

\bibitem[{{Holwerda} {et~al.}(2014){Holwerda}, {Trenti}, {Clarkson}, {Sahu},
  {Bradley}, {Stiavelli}, {Pirzkal}, {De Marchi}, {Andersen}, {Bouwens}, \&
  {Ryan}}]{Holwerda14}
---. 2014, \apj, 788, 77

\bibitem[{Hunter(2007)}]{Hunter07}
Hunter, J.~D. 2007, Computing In Science \& Engineering, 9, 90

\bibitem[{{Jansen} \& {Windhorst}(2018)}]{Jansen18}
{Jansen}, R.~A., \& {Windhorst}, R.~A. 2018, \pasp, 130, 124001

\bibitem[{{Jarrett} {et~al.}(2011){Jarrett}, {Cohen}, {Masci}, {Wright},
  {Stern}, {Benford}, {Blain}, {Carey}, {Cutri}, {Eisenhardt}, {Lonsdale},
  {Mainzer}, {Marsh}, {Padgett}, {Petty}, {Ressler}, {Skrutskie}, {Stanford},
  {Surace}, {Tsai}, {Wheelock}, \& {Yan}}]{Jarrett11}
{Jarrett}, T.~H. {et~al.} 2011, \apj, 735, 112

\bibitem[{{Jones} {et~al.}(2001){Jones}, {Oliphant}, {Peterson}, \&
  Others}]{scipy}
{Jones}, E. {et~al.} 2001, {SciPy}: Open source scientific tools for Python

\bibitem[{{Kelley} {et~al.}(2013){Kelley}, {Fern{\'a}ndez}, {Licandro},
  {Lisse}, {Reach}, {A'Hearn}, {Bauer}, {Campins}, {Fitzsimmons}, {Groussin},
  {Lamy}, {Lowry}, {Meech}, {Pittichov{\'a}}, {Snodgrass}, {Toth}, \&
  {Weaver}}]{Kelley13}
{Kelley}, M.~S. {et~al.} 2013, \icarus, 225, 475

\bibitem[{{Kiss} {et~al.}(2008){Kiss}, {P{\'a}l}, {M{\"u}ller}, \&
  {{\'A}brah{\'a}m}}]{Kiss08}
{Kiss}, C. {et~al.} 2008, \aap, 478, 605

\bibitem[{{Koekemoer} {et~al.}(2011){Koekemoer}, {Faber}, {Ferguson}, {Grogin},
  {Kocevski}, {Koo}, {Lai}, {Lotz}, {Lucas}, {McGrath}, {Ogaz}, {Rajan},
  {Riess}, {Rodney}, {Strolger}, {Casertano}, {Castellano}, {Dahlen},
  {Dickinson}, {Dolch}, {Fontana}, {Giavalisco}, {Grazian}, {Guo}, {Hathi},
  {Huang}, {van der Wel}, {Yan}, {Acquaviva}, {Alexander}, {Almaini}, {Ashby},
  {Barden}, {Bell}, {Bournaud}, {Brown}, {Caputi}, {Cassata}, {Challis},
  {Chary}, {Cheung}, {Cirasuolo}, {Conselice}, {Roshan Cooray}, {Croton},
  {Daddi}, {Dav{\'e}}, {de Mello}, {de Ravel}, {Dekel}, {Donley}, {Dunlop},
  {Dutton}, {Elbaz}, {Fazio}, {Filippenko}, {Finkelstein}, {Frazer}, {Gardner},
  {Garnavich}, {Gawiser}, {Gruetzbauch}, {Hartley}, {H{\"a}ussler},
  {Herrington}, {Hopkins}, {Huang}, {Jha}, {Johnson}, {Kartaltepe},
  {Khostovan}, {Kirshner}, {Lani}, {Lee}, {Li}, {Madau}, {McCarthy},
  {McIntosh}, {McLure}, {McPartland}, {Mobasher}, {Moreira}, {Mortlock},
  {Moustakas}, {Mozena}, {Nandra}, {Newman}, {Nielsen}, {Niemi}, {Noeske},
  {Papovich}, {Pentericci}, {Pope}, {Primack}, {Ravindranath}, {Reddy},
  {Renzini}, {Rix}, {Robaina}, {Rosario}, {Rosati}, {Salimbeni}, {Scarlata},
  {Siana}, {Simard}, {Smidt}, {Snyder}, {Somerville}, {Spinrad}, {Straughn},
  {Telford}, {Teplitz}, {Trump}, {Vargas}, {Villforth}, {Wagner}, {Wandro},
  {Wechsler}, {Weiner}, {Wiklind}, {Wild}, {Wilson}, {Wuyts}, \&
  {Yun}}]{Koekemoer11}
{Koekemoer}, A.~M. {et~al.} 2011, \apjs, 197, 36

\bibitem[{{Kreidberg}(2017)}]{Kreidberg17}
{Kreidberg}, L. 2017, {Exoplanet Atmosphere Measurements from Transmission
  Spectroscopy and Other Planet Star Combined Light Observations}, 100

\bibitem[{{Levan} {et~al.}(2017){Levan}, {Lyman}, {Tanvir}, {Hjorth}, {Mandel},
  {Stanway}, {Steeghs}, {Fruchter}, {Troja}, {Schr{\o}der}, {Wiersema},
  {Bruun}, {Cano}, {Cenko}, {de Ugarte Postigo}, {Evans}, {Fairhurst}, {Fox},
  {Fynbo}, {Gompertz}, {Greiner}, {Im}, {Izzo}, {Jakobsson}, {Kangas},
  {Khandrika}, {Lien}, {Malesani}, {O'Brien}, {Osborne}, {Palazzi}, {Pian},
  {Perley}, {Rosswog}, {Ryan}, {Schulze}, {Sutton}, {Th{\"o}ne}, {Watson}, \&
  {Wijers}}]{Levan17}
{Levan}, A.~J. {et~al.} 2017, \apjl, 848, L28

\bibitem[{{Livermore} {et~al.}(2018){Livermore}, {Trenti}, {Bradley},
  {Bernard}, {Holwerda}, {Mason}, \& {Treu}}]{Livermore18}
{Livermore}, R.~C. {et~al.} 2018, \apjl, 861, L17

\bibitem[{{Malkan} \& {WISP Team}(2013)}]{Malkan13}
{Malkan}, M., \& {WISP Team}. 2013, in Astronomical Society of the Pacific
  Conference Series, Vol. 477, Galaxy Mergers in an Evolving Universe, ed.
  W.-H. {Sun}, C.~K. {Xu}, N.~Z. {Scoville}, \& D.~B. {Sanders}, 255

\bibitem[{Masters {et~al.}(2012)Masters, McCarthy, Burgasser, Hathi, Malkan,
  Ross, Siana, Scarlata, Henry, Colbert, Atek, Rafelski, Teplitz, Bunker, \&
  Dressler}]{Masters12a}
Masters, D. {et~al.} 2012, The Astrophysical Journal Letters, 752, L14

\bibitem[{{Meadows} {et~al.}(2004){Meadows}, {Bhattacharya}, {Reach},
  {Grillmair}, {Noriega-Crespo}, {Ryan}, {Tyler}, {Rebull}, {Giorgini}, \&
  {Elliot}}]{Meadows04}
{Meadows}, V.~S. {et~al.} 2004, \apjs, 154, 469

\bibitem[{{Morishita} {et~al.}(2018){Morishita}, {Abramson}, {Treu}, {Wang},
  {Brammer}, {Kelly}, {Stiavelli}, {Jones}, {Schmidt}, {Trenti}, \&
  {Vulcani}}]{Morishita18}
{Morishita}, T. {et~al.} 2018, ArXiv e-prints

\bibitem[{{Oesch} {et~al.}(2018){Oesch}, {Bouwens}, {Illingworth}, {Labb{\'e}},
  \& {Stefanon}}]{Oesch18a}
{Oesch}, P.~A. {et~al.} 2018, \apj, 855, 105

\bibitem[{{Oesch} {et~al.}(2016){Oesch}, {Brammer}, {van Dokkum},
  {Illingworth}, {Bouwens}, {Labb{\'e}}, {Franx}, {Momcheva}, {Ashby}, {Fazio},
  {Gonzalez}, {Holden}, {Magee}, {Skelton}, {Smit}, {Spitler}, {Trenti}, \&
  {Willner}}]{Oesch16}
---. 2016, \apj, 819, 129

\bibitem[{{Pan} {et~al.}(2017){Pan}, {Kilpatrick}, {Simon}, {Xhakaj},
  {Boutsia}, {Coulter}, {Drout}, {Foley}, {Kasen}, {Morrell},
  {Murguia-Berthier}, {Osip}, {Piro}, {Prochaska}, {Ramirez-Ruiz}, {Rest},
  {Rojas-Bravo}, {Shappee}, \& {Siebert}}]{Pan17}
{Pan}, Y.-C. {et~al.} 2017, \apjl, 848, L30

\bibitem[{{Peek}(2017)}]{Peek17}
{Peek}, K. 2017, Scientific American

\bibitem[{{Petit} {et~al.}(2011){Petit}, {Kavelaars}, {Gladman}, {Jones},
  {Parker}, {Van Laerhoven}, {Nicholson}, {Mars}, {Rousselot}, {Mousis},
  {Marsden}, {Bieryla}, {Taylor}, {Ashby}, {Benavidez}, {Campo Bagatin}, \&
  {Bernabeu}}]{Petit11}
{Petit}, J.-M. {et~al.} 2011, \aj, 142, 131

\bibitem[{{Petit} {et~al.}(2008){Petit}, {Kavelaars}, {Gladman}, {Margot},
  {Nicholson}, {Jones}, {Parker}, {Ashby}, {Campo Bagatin}, {Benavidez},
  {Coffey}, {Rousselot}, {Mousis}, \& {Taylor}}]{Petit08}
{Petit}, J.~M. {et~al.} 2008, Science, 322, 432

\bibitem[{{Pirzkal} {et~al.}(2009){Pirzkal}, {Burgasser}, {Malhotra},
  {Holwerda}, {Sahu}, {Rhoads}, {Xu}, {Bochanski}, {Walsh}, {Windhorst},
  {Hathi}, \& {Cohen}}]{Pirzkal09}
{Pirzkal}, N. {et~al.} 2009, \apj, 695, 1591

\bibitem[{{Pirzkal} {et~al.}(2005){Pirzkal}, {Sahu}, {Burgasser}, {Moustakas},
  {Xu}, {Malhotra}, {Rhoads}, {Koekemoer}, {Nelan}, {Windhorst}, {Panagia},
  {Gronwall}, {Pasquali}, \& {Walsh}}]{Pirzkal05}
---. 2005, \apj, 622, 319

\bibitem[{{Ricker} {et~al.}(2015){Ricker}, {Winn}, {Vanderspek}, {Latham},
  {Bakos}, {Bean}, {Berta-Thompson}, {Brown}, {Buchhave}, {Butler}, {Butler},
  {Chaplin}, {Charbonneau}, {Christensen-Dalsgaard}, {Clampin}, {Deming},
  {Doty}, {De Lee}, {Dressing}, {Dunham}, {Endl}, {Fressin}, {Ge}, {Henning},
  {Holman}, {Howard}, {Ida}, {Jenkins}, {Jernigan}, {Johnson}, {Kaltenegger},
  {Kawai}, {Kjeldsen}, {Laughlin}, {Levine}, {Lin}, {Lissauer}, {MacQueen},
  {Marcy}, {McCullough}, {Morton}, {Narita}, {Paegert}, {Palle}, {Pepe},
  {Pepper}, {Quirrenbach}, {Rinehart}, {Sasselov}, {Sato}, {Seager},
  {Sozzetti}, {Stassun}, {Sullivan}, {Szentgyorgyi}, {Torres}, {Udry}, \&
  {Villasenor}}]{Ricker15}
{Ricker}, G.~R. {et~al.} 2015, Journal of Astronomical Telescopes, Instruments,
  and Systems, 1, 014003

\bibitem[{{Rieke} {et~al.}(2019){Rieke}, {Arribas}, {Bunker}, {Charlot},
  {Finkelstein}, {Maiolino}, {Robertson}, {Willott}, {Windhorst}, \&
  {Eisenstein}}]{Rieke19}
{Rieke}, M. {et~al.} 2019, in \baas, Vol.~51, 45

\bibitem[{{Ryan} {et~al.}(2015){Ryan}, {Mizuno}, {Shenoy}, {Woodward}, {Carey},
  {Noriega-Crespo}, {Kraemer}, \& {Price}}]{Ryan15}
{Ryan}, E.~L. {et~al.} 2015, \aap, 578, A42

\bibitem[{{Ryan} {et~al.}(2011){Ryan}, {Thorman}, {Yan}, {Fan}, {Yan},
  {Mechtley}, {Hathi}, {Cohen}, {Windhorst}, {McCarthy}, \& {Wittman}}]{Ryan11}
{Ryan}, R.~E. {et~al.} 2011, \apj, 739, 83

\bibitem[{{Ryan} {et~al.}(2005){Ryan}, {Hathi}, {Cohen}, \&
  {Windhorst}}]{Ryan05}
{Ryan}, Jr., R.~E. {et~al.} 2005, \apjl, 631, L159

\bibitem[{{Ryan} {et~al.}(2017){Ryan}, {Thorman}, {Schmidt}, {Cohen}, {Hathi},
  {Holwerda}, {Lunine}, {Pirzkal}, {Windhorst}, \& {Young}}]{Ryan17}
---. 2017, \apj, 847, 53

\bibitem[{{Stansberry} {et~al.}(2004){Stansberry}, {Van Cleve}, {Reach},
  {Cruikshank}, {Emery}, {Fernandez}, {Meadows}, {Su}, {Misselt}, {Rieke},
  {Young}, {Werner}, {Engelbracht}, {Gordon}, {Hines}, {Kelly}, {Morrison}, \&
  {Muzerolle}}]{Stansberry04}
{Stansberry}, J.~A. {et~al.} 2004, \apjs, 154, 463

\bibitem[{{Steidel} {et~al.}(1996){Steidel}, {Giavalisco}, {Dickinson}, \&
  {Adelberger}}]{Steidel96}
{Steidel}, C.~C. {et~al.} 1996, \aj, 112, 352

\bibitem[{{Sullivan} {et~al.}(2015){Sullivan}, {Winn}, {Berta-Thompson},
  {Charbonneau}, {Deming}, {Dressing}, {Latham}, {Levine}, {McCullough},
  {Morton}, {Ricker}, {Vanderspek}, \& {Woods}}]{Sullivan15}
{Sullivan}, P.~W. {et~al.} 2015, \apj, 809, 77

\bibitem[{{Trenti}(2014)}]{Trenti14a}
{Trenti}, M. 2014, {Bright Galaxies at Hubble's Detection Frontier: The
  redshift z\~{}9-10 BoRG pure-parallel survey}, HST Proposal

\bibitem[{{Trenti} {et~al.}(2011){Trenti}, {Bradley}, {Stiavelli}, {Oesch},
  {Treu}, {Bouwens}, {Shull}, {MacKenty}, {Carollo}, \&
  {Illingworth}}]{Trenti11}
{Trenti}, M. {et~al.} 2011, \apjl, 727, L39

\bibitem[{{Trenti} {et~al.}(2012){Trenti}, {Bradley}, {Stiavelli}, {Shull},
  {Oesch}, {Bouwens}, {Mu{\~n}oz}, {Romano-Diaz}, {Treu}, {Shlosman}, \&
  {Carollo}}]{Trenti12}
---. 2012, \apj, 746, 55

\bibitem[{{Trenti} \& {Stiavelli}(2008)}]{Trenti08}
{Trenti}, M., \& {Stiavelli}, M. 2008, \apj, 676, 767

\bibitem[{{Trilling} {et~al.}(2016){Trilling}, {Mommert}, {Hora}, {Chesley},
  {Emery}, {Fazio}, {Harris}, {Mueller}, \& {Smith}}]{Trilling16}
{Trilling}, D.~E. {et~al.} 2016, \aj, 152, 172

\bibitem[{{Trilling} {et~al.}(2017){Trilling}, {Valdes}, {Allen}, {James},
  {Fuentes}, {Herrera}, {Axelrod}, \& {Rajagopal}}]{Trilling17}
---. 2017, \aj, 154, 170

\bibitem[{{Trump} {et~al.}(2011){Trump}, {Weiner}, {Scarlata}, {Kocevski},
  {Bell}, {McGrath}, {Koo}, {Faber}, {Laird}, {Mozena}, {Rangel}, {Yan},
  {Yesuf}, {Atek}, {Dickinson}, {Donley}, {Dunlop}, {Ferguson}, {Finkelstein},
  {Grogin}, {Hathi}, {Juneau}, {Kartaltepe}, {Koekemoer}, {Nandra}, {Newman},
  {Rodney}, {Straughn}, \& {Teplitz}}]{Trump11}
{Trump}, J.~R. {et~al.} 2011, \apj, 743, 144

\bibitem[{{van Vledder} {et~al.}(2016){van Vledder}, {van der Vlugt},
  {Holwerda}, {Kenworthy}, {Bouwens}, \& {Trenti}}]{van-Vledder16}
{van Vledder}, I. {et~al.} 2016, \mnras

\bibitem[{{Wadadekar} {et~al.}(2006){Wadadekar}, {Casertano}, {Hook},
  {K{\i}z{\i}ltan}, {Koekemoer}, {Ferguson}, \& {Denchev}}]{Wadadekar06}
{Wadadekar}, Y. {et~al.} 2006, \pasp, 118, 450

\bibitem[{{Williams} {et~al.}(2018){Williams}, {Curtis-Lake}, {Hainline},
  {Chevallard}, {Robertson}, {Charlot}, {Endsley}, {Stark}, {Willmer},
  {Alberts}, {Amorin}, {Arribas}, {Baum}, {Bunker}, {Carniani}, {Crandall},
  {Egami}, {Eisenstein}, {Ferruit}, {Husemann}, {Maseda}, {Maiolino}, {Rawle},
  {Rieke}, {Smit}, {Tacchella}, \& {Willott}}]{Williams18}
{Williams}, C.~C. {et~al.} 2018, \apjs, 236, 33

\bibitem[{{Wright} {et~al.}(2010){Wright}, {Eisenhardt}, {Mainzer}, {Ressler},
  {Cutri}, {Jarrett}, {Kirkpatrick}, {Padgett}, {McMillan}, {Skrutskie},
  {Stanford}, {Cohen}, {Walker}, {Mather}, {Leisawitz}, {Gautier}, {McLean},
  {Benford}, {Lonsdale}, {Blain}, {Mendez}, {Irace}, {Duval}, {Liu}, {Royer},
  {Heinrichsen}, {Howard}, {Shannon}, {Kendall}, {Walsh}, {Larsen}, {Cardon},
  {Schick}, {Schwalm}, {Abid}, {Fabinsky}, {Naes}, \& {Tsai}}]{Wright10}
{Wright}, E.~L. {et~al.} 2010, \aj, 140, 1868

\end{thebibliography}

\begin{table}
\caption{\label{t:ers:usage} The program ID, main instrument, time, science and charged to program, storage volume used, data rate and percentage of onboard storage still available.}
\begin{tabular}{cccccccc}
ProgramID & Instrument & Target & Science & Charge & Volume &  Rate & Available \\

 &  &  & (hours) & (hours) & (GB) & (GB/hr) & (\%) \\
\hline
\hline
1274 & NIRCAM & HD209458 & 6.72 & 10.07 & 56.2 & 8.4 & 3 \\
1185 & NIRCAM & HD189733 & 5.47 & 8.29 & 45.7 & 8.4 & 21 \\
1366 & MIRI & WASP-43 & 24.68 & 29.59 & 43.1 & 1.7 & 26 \\
1274 & NIRCAM & HD189733 & 5.0 & 8.28 & 41.8 & 8.4 & 28 \\
1274 & NIRCAM & HD149026 & 7.51 & 10.56 & 37.8 & 5.0 & 35 \\
1274 & NIRCAM & HD149026 & 7.83 & 10.56 & 36.5 & 4.7 & 37 \\
1185 & NIRCAM & WASP-107 & 7.87 & 10.4 & 34.6 & 4.4 & 40 \\
1185 & NIRCAM & WASP-107 & 7.54 & 10.4 & 34.5 & 4.6 & 40 \\
1274 & NIRCAM & HD209458 & 7.4 & 10.34 & 33.9 & 4.6 & 42 \\
1185 & NIRCAM & GJ436 (x3) & 3.89 & 5.94 & 32.5 & 8.4 & 44 \\
1274 & NIRCAM & WASP77 & 6.78 & 9.28 & 31.9 & 4.7 & 45 \\
1185 & NIRCAM & GJ436 (x3) & 3.56 & 5.94 & 29.8 & 8.4 & 49 \\
1185 & NIRCAM & WASP-80 (x2) & 6.56 & 8.91 & 29.1 & 4.4 & 50 \\
1185 & NIRCAM & WASP-80 (x2) & 6.27 & 8.91 & 29.1 & 4.6 & 50 \\
1224 & NIRSPEC & WASP43 & 22.4 & 28.28 & 24.7 & 1.1 & 57 \\
1201 & NIRISS & WASP121 & 31.54 & 43.62 & 23.8 & 0.8 & 59 \\
1353 & MIRI & WASP-17 (x2) & 9.54 & 12.04 & 16.7 & 1.8 & 71 \\
1185 & NIRCAM & HAT-P-26 & 7.15 & 9.68 & 15.7 & 2.2 & 73 \\
1366 & NIRCAM & WASP-79 & 7.07 & 10.54 & 15.6 & 2.2 & 73 \\
1185 & NIRCAM & HAT-P-26 & 6.75 & 9.68 & 14.9 & 2.2 & 74 \\
1177 & MIRI & HATP19 & 8.19 & 10.58 & 14.5 & 1.8 & 75 \\
1280 & MIRI & WASP107 & 8.1 & 10.36 & 14.2 & 1.8 & 76 \\
1274 & NIRCAM & WASP77 & 6.98 & 9.28 & 13.9 & 2.0 & 76 \\
1177 & MIRI & HATP26 & 7.53 & 9.7 & 13.2 & 1.8 & 77 \\
1279 & MIRI & TRAPPIST1b (x5) & 3.58 & 4.88 & 12.5 & 3.5 & 78 \\
1177 & MIRI & TRAPPIST1b (x5) & 3.57 & 4.87 & 12.4 & 3.5 & 79 \\
1312 & MIRI & HATP26 & 6.98 & 9.07 & 12.3 & 1.8 & 79 \\
1177 & MIRI & WASP-80 (x2) & 6.83 & 8.89 & 12.0 & 1.8 & 79 \\
1281 & MIRI & HATP12 & 6.2 & 8.15 & 10.9 & 1.8 & 81 \\
1353 & NIRSPEC & WASP-17 (x2) & 9.3 & 11.99 & 10.3 & 1.1 & 82 \\
1185 & NIRCAM & HAT-P-19 & 7.91 & 7.8 & 9.4 & 1.2 & 84 \\
1185 & NIRCAM & HAT-P-19 & 8.09 & 10.55 & 9.3 & 1.2 & 84 \\
1366 & NIRSPEC & WASP-79 & 7.93 & 10.59 & 8.9 & 1.1 & 85 \\
1366 & NIRSPEC & WASP-79 & 7.65 & 10.6 & 8.6 & 1.1 & 85 \\
1312 & NIRSPEC & HATP26 (x2) & 6.8 & 9.03 & 7.5 & 1.1 & 87 \\
\hline

\end{tabular}
\end{table}
\setcounter{table}{0}

\begin{table}
\caption{ -- {\em continued}}
\begin{tabular}{cccccccc}
ProgramID & Instrument & Target & Science & Charge & Volume &  Rate & Available \\

 &  &  & (hours) & (hours) & (GB) & (GB/hr) & (\%) \\
\hline
\hline
1177 & MIRI & GJ436 (x2) & 4.28 & 5.93 & 7.5 & 1.8 & 87 \\
1281 & NIRSPEC & HATP12 & 6.16 & 8.57 & 6.9 & 1.1 & 88 \\
1224 & NIRSPEC & WASP107 & 6.16 & 8.51 & 6.9 & 1.1 & 88 \\
1281 & NIRSPEC & HATP12 & 6.16 & 8.32 & 6.8 & 1.1 & 88 \\
1201 & NIRSPEC & WASP107 & 5.99 & 8.3 & 6.7 & 1.1 & 89 \\
1353 & NIRISS & WASP-17 (x2) & 8.67 & 11.98 & 6.5 & 0.8 & 89 \\
1224 & NIRSPEC & GJ3053 & 4.95 & 6.98 & 5.5 & 1.1 & 90 \\
1366 & NIRISS & WASP-79 & 6.59 & 10.48 & 5.0 & 0.8 & 91 \\
1312 & NIRISS & HATP26 & 5.78 & 8.96 & 4.4 & 0.8 & 92 \\
1201 & NIRISS & K2-18 & 5.56 & 8.17 & 4.2 & 0.8 & 93 \\
1201 & NIRISS & WASP107 & 5.17 & 8.12 & 4.0 & 0.8 & 93 \\
1366 & NIRISS & WASP-18 & 5.33 & 8.66 & 3.8 & 0.7 & 93 \\
1331 & NIRSPEC & TRAPPIST-1e (x4) & 3.37 & 5.55 & 3.8 & 1.1 & 93 \\
1201 & NIRISS & HATP1 (x2) & 4.56 & 7.54 & 3.5 & 0.8 & 94 \\
1224 & NIRSPEC & WASP52 & 3.01 & 6.33 & 3.5 & 1.1 & 94 \\
1201 & NIRISS & WASP80 & 4.32 & 7.93 & 3.3 & 0.8 & 94 \\
1201 & NIRISS & LHS1140 (x2) & 4.19 & 6.47 & 3.2 & 0.8 & 94 \\
1201 & NIRISS & GJ3470 & 3.65 & 6.57 & 2.8 & 0.8 & 95 \\
1201 & NIRISS & TRAPPIST1g (x3) & 3.27 & 4.92 & 2.5 & 0.8 & 96 \\
1201 & NIRISS & HD209458 (x2) & 3.08 & 8.08 & 2.4 & 0.8 & 96 \\
1201 & NIRISS & TRAPPIST1f (x4) & 3.08 & 4.69 & 2.3 & 0.8 & 96 \\
1201 & NIRISS & WASP69 & 2.69 & 7.16 & 2.2 & 0.8 & 96 \\
1201 & NIRSPEC & TRAPPIST1d (x2) & 1.82 & 4.2 & 2.1 & 1.2 & 96 \\
1201 & NIRISS & GJ1132 (x4) & 2.28 & 4.1 & 1.8 & 0.8 & 97 \\
1201 & NIRISS & GJ436 & 1.6 & 4.64 & 1.3 & 0.8 & 98 \\
1201 & NIRISS & HATP1 (x2) & 0.25 & 0.87 & 0.2 & 1.0 & 100 \\
1201 & NIRISS & HD209458 (x2) & 0.22 & 0.84 & 0.2 & 1.0 & 100 \\
\hline

\end{tabular}
\end{table}
\end{document}